\documentclass[aps,showkeys,11pt,twoside]{revtex4}

\usepackage[english]{babel}
\usepackage{graphicx,epic,eepic,epsfig,amsmath,latexsym,amssymb,verbatim}

\usepackage{theorem}
\newtheorem{definition}{Definition}
\newtheorem{proposition}[definition]{Proposition}
\newtheorem{example}[definition]{Example}

\newtheorem{theorem}[definition]{Theorem}
\newtheorem{corollary}[definition]{Corollary}

\def\squareforqed{\hbox{\rlap{$\sqcap$}$\sqcup$}}
\def\qed{\ifmmode\squareforqed\else{\unskip\nobreak\hfil
\penalty50\hskip1em\null\nobreak\hfil\squareforqed
\parfillskip=0pt\finalhyphendemerits=0\endgraf}\fi}
\def\endenv{\ifmmode\;\else{\unskip\nobreak\hfil
\penalty50\hskip1em\null\nobreak\hfil\;
\parfillskip=0pt\finalhyphendemerits=0\endgraf}\fi}
\newenvironment{proof}{\noindent \textbf{{Proof~} }}{\qed}

\mathchardef\ordinarycolon\mathcode`\: \mathcode`\:=\string"8000
\def\vcentcolon{\mathrel{\mathop\ordinarycolon}}
\begingroup \catcode`\:=\active
  \lowercase{\endgroup
  \let :\vcentcolon
  }

\newcommand{\nc}{\newcommand}
\nc{\rnc}{\renewcommand} \nc{\beq}{\begin{equation}}
\nc{\eeq}{{\end{equation}}} \nc{\beqa}{\begin{eqnarray}}
\nc{\eeqa}{\end{eqnarray}} \nc{\lbar}[1]{\overline{#1}}
\nc{\bra}[1]{\langle#1|} \nc{\ket}[1]{|#1\rangle}
\nc{\ketbra}[2]{|#1\rangle\!\langle#2|}
\nc{\braket}[2]{\langle#1|#2\rangle}
\nc{\proj}[1]{|#1\rangle\!\langle #1 |}
\nc{\avg}[1]{\langle#1\rangle} \rnc{\max}{\operatorname{max}}
\nc{\Rank}{\operatorname{Rank}}
\nc{\smfrac}[2]{\mbox{$\frac{#1}{#2}$}}
\nc{\tr}{\operatorname{Tr}} \nc{\ox}{\otimes} \nc{\dg}{\dagger}
\nc{\dn}{\downarrow} \nc{\cA}{{\cal A}} \nc{\cB}{{\cal B}}
\nc{\cC}{{\cal C}} \nc{\cD}{{\cal D}} \nc{\cE}{{\cal E}}
\nc{\cF}{{\cal F}} \nc{\cG}{{\cal G}} \nc{\cH}{{\cal H}}
\nc{\cI}{{\cal I}} \nc{\cJ}{{\cal J}} \nc{\cK}{{\cal K}}
\nc{\cL}{{\cal L}} \nc{\cM}{{\cal M}} \nc{\cN}{{\cal N}}
\nc{\cO}{{\cal O}} \nc{\cP}{{\cal P}} \nc{\cR}{{\cal R}}
\nc{\cS}{{\cal S}} \nc{\cT}{{\cal T}} \nc{\rU}{{\cal U}}
\nc{\cX}{{\cal X}} \nc{\cZ}{{\cal Z}}
\nc{\csupp}{{\operatorname{csupp}}}
\nc{\qsupp}{{\operatorname{qsupp}}} \nc{\var}{\operatorname{var}}
\nc{\rar}{\rightarrow} \nc{\lrar}{\longrightarrow}
\nc{\polylog}{\operatorname{polylog}}
\nc{\supp}{\operatorname{supp}\,} \nc{\1}{\openone}

\nc{\RR}{{{\mathbb R}}} \nc{\CC}{{{\mathbb C}}} \nc{\FF}{{{\mathbb
F}}} \nc{\NN}{{{\mathbb N}}} \nc{\ZZ}{{{\mathbb Z}}}
\nc{\PP}{{{\mathbb P}}} \nc{\QQ}{{{\mathbb Q}}} \nc{\UU}{{{\mathbb
U}}} \nc{\WW}{{{\mathbb W}}} \nc{\EE}{{{\mathbb E}}}
\nc{\id}{{\operatorname{id}}}

\nc{\ob}[1]{#1}

\begin{document}

\title{Robustness of quantum Markov chains}

\date{5th November 2006}

\author{Ben Ibinson, Noah Linden and Andreas Winter}
\affiliation{Department of Mathematics, University of Bristol, Bristol BS8 1TW, U.~K.
             \protect\\ Email: \{ben.ibinson,\,{}n.linden,\,{}a.j.winter\}@bristol.ac.uk}

\begin{abstract}
If the conditional information of a classical probability distribution of 
three random variables is zero, then it obeys a Markov chain condition. 
If the conditional information is close to zero, then it is known that 
the distance (minimum relative entropy) of the distribution to the nearest 
Markov chain distribution is precisely the conditional information.   
We prove here that this simple situation does not obtain for quantum 
conditional information.   We show that for tri-partite quantum states the 
quantum conditional information is always a lower bound for the minimum 
relative entropy distance to a quantum Markov chain state, but the distance
can be much greater; indeed the two quantities can be of different asymptotic 
order and may even differ by a dimensional factor.
\end{abstract}


\maketitle

\section{Introduction}
\label{sec:intro} From the point of view of information theory, as
well as physics, it is very interesting to know when entropy or,
more generally, information inequalities are saturated. For
example, the basic quantities von Neumann entropy $S(A) =
S(\rho_A) = -\tr \rho_A\log\rho_A$, quantum mutual information
$I(A:B)=S(A)+S(B)-S(AB)$ for a bipartite state $\rho_{AB}$ and
conditional mutual information $I(A:C|B)=S(AB)+S(BC)-S(B)-S(ABC)$
for a tripartite state $\rho_{ABC}$ are all non-negative; for the
latter two this is known as the subadditivity and strong
subadditivity of the entropy, respectively~\cite{Lieb:Ruskai}. The
entropy is $0$ if and only if the state is pure, and the mutual
information is $0$ if and only if the state $\rho_{AB}$ is a
product state, $\rho_{AB} = \rho_A \ox \rho_B$.

However, in many applications it is not the case or not known that
the state is exactly pure or a product, only that it is very close
to being so. In such situations, there are continuity bounds on
entropic quantities that one can use to quantify how small the
entropy or mutual information is. Fannes' inequality~\cite{fannes}
states that if $\| \rho_A - \sigma_A \|_1 \leq \epsilon \leq 1/e$
(with the trace norm $\| X \|_1 := \tr |X| = \tr \sqrt{X^* X}$),
then
\begin{equation}
  | S(\rho)-S(\sigma) | \leq -\epsilon\log\epsilon + \epsilon \log
  d_A,\label{fannes-ineq}
\end{equation}
where $d_A$ is the dimension of the Hilbert space supporting the
states. (``$\log$'' in this paper is always the binary logarithm;
the natural logarithm is denoted ``$\ln$''.) In particular, if
$\rho$ has trace distance $\epsilon\leq 1/e$ to a pure state, then
$S(\rho) \leq -\epsilon\log\epsilon + \epsilon \log d_A$.
Recently, Alicki and Fannes~\cite{AlickiFannes04} proved an
extension of the Fannes inequality to quantum conditional entropy
$S(A|B) = S(AB)-S(B)$ for bipartite states $\rho_{AB}$ and
$\sigma_{AB}$: if $\| \rho_A - \sigma_A \|_1 \leq \epsilon \leq
1$, then
\begin{equation}
  | S(A|B)_\rho - S(A|B)_\sigma | \leq -2\epsilon\log\epsilon
                                       -2(1-\epsilon)\log(1-\epsilon)
                                       +4\epsilon \log d_A.\label{alicki-fannes-ineq}
\end{equation}
The crucial observation here is that the bound only depends on
$\epsilon$ and $d_A$, not $d_B$ as the bound yielded by a naive
application of the original Fannes inequality. This gives an upper
bound on the mutual information for a state that is at trace
distance $\epsilon$ from a product state (using convexity of the
trace distance, and (\ref{fannes-ineq}) and
(\ref{alicki-fannes-ineq}) together with the triangle inequality).

Conversely, one may ask, if say the entropy of a state is small,
$S(\rho) \leq \epsilon$, is it close to being pure? Indeed yes, as
the following argument shows. Fix a diagonalisation of $\rho$,
$\rho = \sum_{i=1}^{d_A} \lambda_i \proj{e_i}$ with eigenvalues
$\lambda_i$ arranged in decreasing order. Then, as $-x\log x \geq
x$ for $0\leq x\leq 1/2$,
\begin{equation}
  \epsilon \geq S(\rho) =
    \sum_{i=1}^{d_A} -\lambda_i \log\lambda_i \geq \sum_{i=2}^{d_A} \lambda_i = 1-\lambda_1.
\end{equation} Hence,
\begin{equation}
  \| \rho - \proj{e_1} \|_1 = 2(1-\lambda_1) \leq 2\epsilon.
\end{equation}
Note however that this bound and Fannes' inequality are not
``inverse'' to each other; plugging the $2\epsilon$ into the
Fannes bound yields something much larger than order $\epsilon$.

Similarly, what can we say about the state when $I(A:B) \leq
\epsilon$? Here, a new quantity, the relative
entropy $D(\rho\|\sigma) = \tr\rho(\log\rho-\log\sigma)$, comes
into play, when we observe that $I(A:B)_\rho =
D(\rho_{AB}\|\rho_A\ox\rho_B)$. Invoking another inequality
between distance measures for states, namely Pinsker's inequality,
see~\cite{Fuchs:vandeGraaf},
\begin{equation}
  D(\rho\|\sigma) \geq \left( \frac{1}{2\ln 2}\| \rho-\sigma \|_1 \right)^2,
\end{equation}
we conclude that $\| \rho_{AB} - \rho_A\ox\rho_B \|_1 \leq
2\sqrt{\epsilon}$. Note that in both examples discussed, we found
an explicit candidate for the closest pure/product state to the
given state (as can be checked), and that the bound on the trace
distance depends only on $\epsilon$, not on dimensions as in the
converse Fannes-style inequalities. Third, that the relative
entropy gives even tighter control on the distance due to
Pinsker's inequality.

\medskip
In this paper we study the quantum conditional information. If the
quantum conditional information of a tri-partite state $\rho$
vanishes, then $\rho$ obeys a quantum Markov chain condition. Here
we analyze what can be said if $\rho$ has small quantum
conditional information; in particular we investigate how close it
is to a Markov chain state.  The motivation is partly classical
(e.g.~cryptographic~\cite{MHorodecki}), but in the quantum case a
strong motivation comes from considerations of new entropy
inequalities: in~\cite{Linden:Winter} a so-called constrained
inequality for the von Neumann entropies of subsystems was found,
namely a relation which is valid provided three quantum
conditional mutual informations are zero. The desire to turn this
into an unconstrained, universal inequality lead to speculations
that if one understood the near-vanishing of these constraints,
then perhaps a trade-off between the constraints and the new
inequality solely in terms of entropies might be established.

In section~\ref{sec:classical} we
review, as a model, the classical case, where it turns out that
the conditional mutual information is exactly the minimum relative
entropy distance between the distribution and the closest Markov
chain distribution. In section~\ref{sec:quantum} we formulate the
analogous quantum problem, which we analyse in the rest of the
paper: section~\ref{sec:partial-results} presents several
simplifications of the question -- we prove continuity of the
minimum relative entropy, and that it is lower bounded by the
quantum conditional mutual information, and some useful formulas
for later numerical and analytical evaluation of the quantity.
Then, in section~\ref{sec:pure}, we specialise to pure states: we
relate the minimum relative entropy to the so-called entanglement
of purification, and for a large family of states show upper and
lower bounds of matching order. These results are then used in
section~\ref{sec:examples} to provide examples of states for which
the minimum relative entropy is much larger than the quantum
conditional mutual information, and also ones where the dimension
enters explicitly, showing that the classical and the quantum case
are very different indeed.

\section{Classical case}
\label{sec:classical} In the classical case, Markov chain
distributions are used to define the conditional mutual information.
A classical distribution $P_{X_1X_2\ldots X_n}(x_1,x_2,\ldots, x_n)$
form a Markov chain denoted as $X_1 \rightarrow X_2\rightarrow X_3
\rightarrow \ldots\rightarrow X_n$ if the distribution can be
written as
\begin{equation}
P_{X_1 X_2\ldots X_n} (x_1,x_2,\ldots ,x_3)
  = P_{X_1 X_2}(x_1,x_2)P_{X_3|X_2}(x_3|x_2) \ldots P_{X_n|X_{n-1}}(x_n|x_{n-1})
\end{equation}
If we take any three linked variables in the above Markov chain i.e.
$X_{\alpha-1}$,$ X_\alpha$,$X_{\alpha+1}$ then the conditional
distribution of $P_{X_{\alpha+1}|X_\alpha\ldots X_1}(x_{\alpha+1}|x_\alpha\ldots x_1)$
depends only on $X_\alpha$, and $X_{\alpha+1}$ is conditionally independent of
$X_{\alpha-1}$, given $X_\alpha$.
Consider three random variables $X,Y$ and $Z$ which form a Markov chain
$X \rightarrow Y \rightarrow Z$. The probability distribution for
this system is
\begin{equation}
\label{eq:markovclass}
\begin{split}
P_{XYZ}(xyz)&=P_{XY}(xy)P_{Z|Y}(z|y)\\
            &=P_Y(y)P_{X|Y}(x|y)P_{Z|Y}(z|y).
\end{split}
\end{equation}
Aside, we define the conditional mutual information as,
\begin{equation}
I(X:Z|Y)=\sum_{x,y,z}
P_{XYZ}(xyz)\log{\frac{P_{XZ|Y}(xz|y)}{P_{X|Y}(x|y)P_{Z|Y}(z|y)}}
\end{equation}
Note that throughout this section we use the convention $0 \log 0 =0$.
(This is justified by looking at the behavior of $x\log{x}$ as $x \rightarrow0$.)
This conditional mutual information is
equal to zero if and only if for all $x$, $y$ and $z$,
\begin{equation}
\frac{P_{XZ|Y}(xz|y)}{P_{X|Y}(x|y)P_{Z|Y}(z|y)}=1
\end{equation}
Therefore,
\begin{equation}
\begin{split}
{P_{XZ|Y}(xz|y)}&={P_{X|Y}(x|y)P_{Z|Y}(z|y)}\\
{P_{XYZ}(xyz)}&={P_Y(y)P_{X|Y}(x|y)P_{Z|Y}(z|y)}.
\end{split}
\end{equation}
Hence a classical Markov chain distribution is characterized by zero
conditional mutual information. The classical case is characterized
by an exact correspondence between the conditional mutual
information and the relative entropy distance to the set of Markov
chains: for any joint distribution $P_{XYZ}$ of three random
variables $X,Y,Z$~\cite{MHorodecki},
\begin{equation}
  I(X:Z|Y) = \min\bigl\{ D(P\|Q) : Q \text{ Markov} \bigr\}.
\end{equation}
It can be shown that the Markov chain required to minimise this
quantity is
\begin{equation}
\label{eq:c-structure}
  Q_{XYZ}(xyz)= P_Y(y) P_{X|Y}(x|y) P_{Z|X}(z|x).
\end{equation}
\begin{proof}
Imagine a joint probability distribution $Q_{XYZ}$ that also forms a
general Markov chain:
\begin{equation}
  Q_{XYZ}(xyz)= Q_Y(y) Q_{X|Y}(x|y) Q_{Z|Y}(z|y).
\end{equation}
We can write the probability distribution of $P_{XYZ}$ as follows
\begin{equation}
P_{XYZ}(xyz)=P_{XYZ}(xyz)=P_{Y}(y)P_{Z|Y}(z|y)P_{X|YZ}(x|yz),
\end{equation}
therefore the relative entropy between the two probability
distributions is
\begin{equation}
  D(P\|Q)=\sum_{xyz} P_{XYZ}(xyz) \log
  \frac{P_Y(y)P_{Z|Y}(z|y)P_{X|YZ}(x|yz)}{Q_Y(y)Q_{Z|Y}(z|y)Q_{X|Y}(x|y)}.
\end{equation}
Since we have a product of logarithms we can represent the relative
entropy as such,
\begin{equation}
\label{eq:logsum}
  D(P\|Q)=\sum_{xyz} P_{XYZ}(xyz) \bigg(\log
  \frac{P_Y(y)}{Q_Y(y)} + \log
  \frac{P_{Z|Y}(z|y)}{Q_{Z|Y}(z|y)} + \log
  \frac{P_{X|YZ}(x|yz)}{Q_{X|Y}(x|y)}\bigg).
\end{equation}
On inspection of the final term we can use the following equivalence
\begin{equation}
  \frac{P_{X|YZ}(x|yz)}{Q_{X|Y}(x|y)}=\frac{P_{XYZ}(xyz)}{P_{YZ}(yz)Q_{X|Y}(x|y)} =
  \frac{P_{Z|XY}(z|xy)}{P_{Z|Y}(z|y)}\frac{P_{X|Y}(x|y)}{Q_{X|Y}(x|y)}.
\end{equation}
Observing that the first two terms of eq. (\ref{eq:logsum}) are
relative entropy terms, we have
\begin{equation}
\begin{split}
\label{eq:rellog}
  D(P\|Q) = D\big(P_Y(y)\|&Q_Y(y)\big)+D\big(P_{Z|Y}(z|y)\|Q_{Z|Y}(z|y)\big) \\
                          &+ D\big(P_{X|Y}(x|y)\|Q_{X|Y}(x|y)\big)+
                             \sum_{xyz} P_{XYZ}(xyz) \log \frac{P_{Z|XY}(z|xy)}{P_{Z|Y}(z|y)}.
\end{split}
\end{equation}
Note that the only terms that depend on the
distribution of $Q$ are the first three relative entropy terms.
Since relative entropy is non-negative and $D(S\|T)=0$ if and only if $S=T$,
the Markov chain that provides the minimum relative entropy
between $P$ and $Q$ can achieved by setting these terms to zero which
gives the required distribution in (\ref{eq:c-structure}). This
concludes the proof.
\end{proof}
From this result it is simple to show that conditional mutual
information can be achieved. Since we know the relative entropy
terms in eq. (\ref{eq:rellog}) are zero if we use $Q$ as the
minimising Markov chain:
\begin{equation}
\label{eq:condmut}
    D(P\|Q) = \sum_{xyz} P_{XYZ}(xyz) \log \frac{P_{Z|XY}(z|xy)}{P_{Z|Y}(z|y)}.
\end{equation}
Using the following equivalence
\begin{equation}
P_{Z|XY}(z|xy) = \frac{P_{XYZ}(xyz)}{P_{XY}(xy)} =
\frac{P_{XYZ}(xyz)}{P_Y(y)} \frac{P_Y(y)}{P_{XY}(xy)} =
\frac{P_{XZ|Y}(xz|y)}{P_{X|Y}(x|y)}.
\end{equation}
We can substitute this into (\ref{eq:condmut}) to produce the
final result
\begin{equation}
    D(P\|Q) = \sum_{xyz} P_{XYZ}(xyz) \log \frac{P_{XZ|Y}(xz|y)}{P_{X|Y}(x|y)P_{Z|Y}(z|y)} =
    I(X:Z|Y).
\end{equation}

\section{Quantum analogue}
\label{sec:quantum}
A quantum analogue of (short) Markov chains,
i.e.~quantum states of some tripartite system $ABC$ with a
suitably defined Markov property, was first proposed by Accardi
and Frigerio~\cite{Acc-Frig}. In finite Hilbert space dimension,
which will be the case we will consider in the present paper, this
property reads as follows: $\mu_{ABC}$ is a quantum Markov state
if there exists a quantum channel, i.e.~a completely positive and
trace preserving (c.p.t.p.) map
$T:{\cal B}(B) \longrightarrow {\cal B}(B) \ox {\cal B}(C)$ such that
$\mu_{ABC} = (\id_A \otimes T)\mu_{AB}$, with
$\mu_{AB} = \tr_C \mu_{ABC}$. In~\cite{Petz} it was shown that this Markov
condition is equivalent to vanishing conditional mutual information,
\begin{equation}
  I(A:C|B)_\mu = 0,
\end{equation}
just as in the classical case, and in~\cite{q-Markov} the most
general form of such states was given, as follows: system $B$ has
a direct sum decomposition into tensor products,
\begin{equation}
  \label{eq:B-structure}
  B = \bigoplus_{j} {b_j^L} \ox {b_j^R},
\end{equation}
such that
\begin{equation}
  \label{eq:q-structure}
  \mu_{ABC} = \bigoplus_{j} p_j \mu^{(j)}_{A b_j^L} \ox \mu^{(j)}_{b_j^R C}.
\end{equation}
Note that this precisely generalises
eq.~(\ref{eq:markovclass}). We introduce the notation $\delta$
for the direct sum decomposition of $B$. Note that we can always think
of $\mathcal{H}_B$ as being a subspace of a larger Hilbert space
$\mathcal{H}_{\widehat{B}}$ (for which inclusion we use the shorthand
$B\hookrightarrow\widehat{B}$). This doesn't change the fact
that a state is a Markov chain state or not, but it leads to
more possibilities of decomposing the ambient Hilbert space as a sum
of products as in eq.~(\ref{eq:B-structure}). In other words, in a larger
space there is a larger set of quantum Markov chains. This latter is evidently
going to be relevant when comparing a given state $\rho_{ABC}$ to
the class of Markov chain states: in general we will have to admit
that all three systems $A$, $B$, $C$ are subspaces of larger
Hilbert spaces, and we have to take into account the Markov states on the
extended system.

Now we go on to develop some formalism to deal with these embeddings:
If we have the embedded system $B \hookrightarrow \widehat{B}$ then we define
$\delta \ \equiv\ \Bigl( B \hookrightarrow \widehat{B} = \bigoplus_j B_j \Bigr)$
as both the isometric embedding and the orthogonal decomposition of the
embedding system. For a specific such direct sum decomposition $\delta$
we introduce $\tau=\tau_\delta$ for the family of tensor product
decompositions of the $B_j$, which are, w.l.o.g., embeddings
$\tau_\delta \ \equiv\  \Bigl( \tau_j : B_j \hookrightarrow b_j^L
\otimes b_j^R \Bigr)_j$. Note that this latter only gives us increased
flexibility: we could as well demand that each $\tau_j$ is actually a
unitary isomorphism between $B_j$ and $b_j^L \otimes b_j^R$, because
one can always blow up the spaces $B_j$, extending the isometry to a
unitary.

This brings us to the main question of this paper: for given state
$\rho_{ABC}$, to find
\begin{equation}
  \label{eq:Delta}
  \Delta(\rho) := \inf\bigl\{ D(\rho\|\mu) : \mu \text{ Markov} \bigr\},
\end{equation}
and to compare it to $I(A:C|B)_\rho$. The remainder of this paper
will be devoted to a study of the properties of this function. To be
precise, we would like to consider $\mu$ to be a Markov state on
a tripartite system $\widehat{A}\widehat{B}\widehat{C}$, with $A
\subset \widehat{A}$, $B \subset \widehat{B}$ and $C \subset
\widehat{C}$ (with $\rho$ understood to be also a state on
$\widehat{A}\widehat{B}\widehat{C}$ via these embedding), which is
why above we have to use the infimum, since the dimension of
$\widehat{A}\widehat{B}\widehat{C}$ is unbounded. This appears to be
necessary for the reason that the decompositions as in
eq.~(\ref{eq:q-structure}) depend on the dimension of $B$.

We will show below (in the next section) that
w.l.o.g.~$A=\widehat{A}$ and $C=\widehat{C}$, and $\dim
\widehat{B} \leq d_B^4$, so that the infimum is actually a
minimum. We also show lower bounds on $\Delta$ comparing it to
$I(A:C|B)$, in particular exhibiting examples of states $\rho$
with $\Delta(\rho) \gg I(A:C|B)_\rho$.

\section{General Properties of $\Delta$}
\label{sec:partial-results} Here we show that the problem of
determining the minimum relative entropy to a Markov state is
really only a minimisation over decompositions of the type
(\ref{eq:B-structure}) for $\widehat{B}$.
For given dimensions of the quantum
system, there is only a finite number number of decomposition
types. Therefore we need to perform a finite-dimensional
optimisation for each decomposition (some of which we can perform
explicitly) and choose the global minimum.

\begin{proposition}
\label{prop:given-delta-tau} The optimal state for given direct
sum and tensor decomposition we denote $\omega[\delta,\tau]$
describing the specific direct sum as $\delta$ and the chosen
tensor decomposition for that direct sum $\tau=\tau_\delta$. We
obtain $\omega[\delta,\tau]$ by the following procedure: first,
with the subspace projections $P_j$ onto
$b_j^L \ox b_j^R \subset \widehat{B}$, let
\begin{equation}
  \label{eq:omega-oplus}
  \omega[\delta]:= \bigoplus_j (\1_{AC}\ox P_j) \rho (\1_{AC}\ox P_j)
                 = \bigoplus_j q_j \omega^{(j)}_{A b_j^L b_j^R C},
\end{equation}
where for each part $j$ of the direct sum for the given
decomposition, we project system $\widehat{B}$ via the corresponding
projections $P_j$ to produce $\omega^{(j)}_{Ab_j^Lb_j^RC}$ with
corresponding probability $q_j$. Then, form the reduced states
$\sigma^{(j)}_{A b_j^L} = \tr_{b_j^R C} \omega^{(j)}_{A b_j^L
b_j^R C}$ and $\chi^{(j)}_{b_j^R C} = \tr_{A b_j^L}
\omega^{(j)}_{A b_j^L b_j^R C}$, and let
\begin{equation}
  \label{eq:omega-oplus-ox}
  \omega[\delta,\tau] := \bigoplus_j q_j \sigma^{(j)}_{A b_j^L} \ox \chi^{(j)}_{b_j^R C}.
\end{equation}
With these definitions, it is easy to work out that
\begin{equation}
  \label{eq:relent-oplus-ox}
  D\bigl( \rho\|\omega[\delta,\tau] \bigr)
      = -S(\rho) + H(\underline{q}) + \sum_j q_j\Bigl( S\bigl(\sigma^{(j)}_{A b_j^L} \bigr)
                                                  + S\bigl(\chi^{(j)}_{b_j^R C} \bigr)
                                                  \Bigr).
\end{equation}
Then, among all Markov states with decomposition
(\ref{eq:B-structure}) of $\widehat{B}$, $\omega[\delta,\tau]$ is the
one with smallest relative entropy, given by eq.~(\ref{eq:relent-oplus-ox}).
\end{proposition}
\begin{proof}
The relative entropy between a general state $\rho_{ABC}$ and a
general quantum Markov state $\mu_{ABC}$ is, with given decompositions
$\delta$ and $\tau$,
\begin{equation}
  D(\rho_{ABC}\|\mu_{ABC}) = -S(\rho_{ABC})-\tr(\rho_{ABC} \log \mu_{ABC}),
\end{equation}
where the general quantum Markov state is defined as
\begin{equation}
\mu_{ABC} = \bigoplus_j p_j \mu^{(j)}_{Ab_j^L} \otimes \mu^{(j)}_{b_j^RC}.
\end{equation}
Therefore we can calculate the logarithm of $\mu_{ABC}$,
\begin{equation}
  \log \mu_{ABC} = \bigoplus_j \bigg( \log p_j(\1_{AC} \otimes P_j) +
  \log (\mu^{(j)}_{Ab_j^L} \otimes \mu^{(j)}_{b_j^RC})\bigg) =:
  \bigoplus_j L_j,
\end{equation}
where $P_j$ are the subspace projections from $\widehat{B}$ onto $b_j^Lb_j^R$.
For clarity we assume $\rho$ indicates the state over all parties
unless otherwise indicated.
\begin{equation}
\tr \rho_{ABC} \log \mu_{ABC} = \sum_j \tr \bigg((\1_{AC} \otimes
  P_j)\rho(\1_{AC} \otimes P_j) L_j \bigg).
\end{equation}
Now $(\1_{AC} \otimes P_j)\rho(\1_{AC} \otimes P_j) =
q_j\omega^{(j)}_{Ab_j^Lb_j^RC}$ with $\tr
\omega^{(j)}_{Ab_j^Lb_j^RC} = 1$. Therefore,
\begin{align}
\tr \rho_{ABC} \log \mu_{ABC}& = \sum_j \tr q_j
\omega^{(j)}_{Ab_j^Lb_j^RC} \log p_j(\1_{AC} \otimes P_j) + \sum_j
\tr q_j \omega^{(j)}_{Ab_j^Lb_j^RC} \log (\mu^{(j)}_{Ab_j^L}
\otimes \mu^{(j)}_{b_j^RC})\\
& =\sum_j q_j \log p_j + \sum_j q_j \tr \omega^{(j)}_{Ab_j^Lb_j^RC}
\log(\mu^{(j)}_{Ab_j^L} \otimes \mu^{(j)}_{b_j^RC})\\
&= -H(\underline{q})-D(\underline{q}\|\underline{p})\nonumber\\
   &\quad -\sum_j q_j
\bigg(S(\sigma^{(j)}_{Ab_j^L})+S(\chi^{(j)}_{b_j^RC})
+D(\sigma^{(j)}_{Ab_j^L}\|\mu^{(j)}_{Ab_j^L})+D(\chi^{(j)}_{b_j^RC}
\|\mu^{(j)}_{b_j^RC})\bigg),
\end{align}
where $\sigma^{(j)}_{Ab_j^L}= \tr_{b_j^RC}
\omega^{(j)}_{Ab_j^Lb_j^RC}$ and $\chi^{(j)}_{b_j^RC}=
\tr_{Ab_j^L} \omega^{(j)}_{Ab_j^Lb_j^RC}$. For a given
decomposition of system $\widehat{B}$, the subspace projections $P_j$ and
hence $q_j$ and $\omega^{(j)}_{Ab_j^Lb_j^RC}$ are fixed. Since we
want to minimise the relative entropy we want to maximise the
quantity $\tr \rho_{ABC} \log \mu_{ABC}$. Therefore to maximise
the first relative entropy term we set $p_j=q_j$. For the sum we
consider each $i$ individually and only have freedom of setting
the last two relative entropy terms to zero. Therefore we can
maximise this expression by setting
$\mu^{(j)}_{Ab_j^L}=\sigma^{(j)}_{Ab_j^L}$ and
$\mu^{(j)}_{b_j^RC}=\chi^{(j)}_{b_j^RC}$. This concludes the
proof.
\end{proof}

A nice observation is that the relative entropy of interest can be
decomposed into two relative entropies, as follows:
\begin{equation}
  \label{eq:2-relents}\begin{split}
  D\bigl( \rho\|\omega[\delta,\tau]\bigr)
     & = D\bigl( \rho\|\omega[\delta]\bigr)
         + D\bigl( \omega[\delta]\|\omega[\delta,\tau] \bigr)\\
      &
      = D\bigl( \rho\|\omega[\delta]\bigr)
         + \sum_j q_j I(A b_j^L : b_j^R C)_{\omega^{(j)}}.
         \end{split}
\end{equation}
Note that this result has an important consequence for the infimum
defining $\Delta(\rho)$: we only need to worry about embedding $B$
into a larger system $\widehat{B}$; $A$ and $C$ can, w.l.o.g.,
stay the same.

\begin{theorem}
  \label{prop:finite-min}
  The infimum of eq.~(\ref{eq:Delta}) is achieved on a decomposition
  $\widehat{B} = \bigoplus_{i=1}^{k \leq d_B^2} b_j^L \otimes b_j^R$,
  with $\dim b_j^L, \dim b_j^R \leq d_B$. In particular, because
  it is one of a continuous function over a compact domain, the infimum
  is actually a minimum. Also, it means that
  $\Delta(\rho)$, as the minimum of a continuous function over a
  compact domain, is itself a continuous function of its
  argument $\rho$.
\end{theorem}

\noindent
The reader may wish to skip the rather lengthy and somewhat technical proof
of this theorem; note however that in it some notation is introduced
which is referred to later.
\medskip

\begin{proof}
 The proof has two parts -- first, that the direct sum decomposition $\delta$
 may be taken to have only $d_B^2$ terms, and second, that each direct
 summand may be embedded into a space of not more than $d_B\times d_B$ dimensions. These
 two arguments are quite independent of each other; we start with the first.

 {\bf 1.}
 For given embedding $B\hookrightarrow \widehat{B}$ and decomposition
 of $\widehat{B}$, we have
 \begin{equation}
   \omega[\delta] = \bigoplus_j (\1_{AC}\otimes P_j)\rho_{ABC}(\1_{AC}\otimes P_j)
                  = \bigoplus_j (\1_{AC}\otimes P_jP)\rho_{ABC}(\1_{AC}\otimes PP_j),
 \end{equation}
 with the projector $P$ of $\widehat{B}$ onto $B$. Note that the operators
 $P_jP$ form a complete Kraus system:
 \begin{equation}
   \sum_j (P_jP)^\dagger (P_jP) = \sum_j PP_jP = P = \1_B.
 \end{equation}
 Hence the operators $M_j=PP_jP$ form a POVM on $B$, and introducing an
 auxiliary register $J$ with orthogonal states $\ket{j}$ to reflect the
 direct sum, $\omega[\delta]$ is equivalent, up to local isometries,
 to the state
 \begin{equation}
   \Omega = \sum_j \bigl( \1_{AC}\otimes\sqrt{M_j} \bigr)
                      \rho_{ABC}
                   \bigl( \1_{AC}\otimes\sqrt{M_j} \bigr) \otimes \proj{j}_J
 \end{equation}
 At the same time, the embedding $\tau_j$ can be reinterpreted as a
 family of isometries
 $\tau_j: B \hookrightarrow b^L \otimes b^R$ controlled by the content $j$
 of the $J$-register (note that we may, w.l.o.g., assume that the $\tau_j$
 all map into the same tensor product space), so that the state after
 the action of $\tau$ is
 \begin{equation}
   \label{eq:Omega}
   \Omega_{Ab^Lb^RCJ} = \sum_j \bigl( \1_{AC}\otimes\tau_j\sqrt{M_j} \bigr)
                                  \rho_{ABC}
                               \bigl( \1_{AC}\otimes\sqrt{M_j}\tau_j^\dagger \bigr)
                                                                  \otimes \proj{j}_J.
 \end{equation}
 In this notation, our formula (\ref{eq:relent-oplus-ox}) can be rewritten as
 \begin{equation}
   \label{eq:D-compact}
   D(\rho\|\omega[\delta,\tau]) = -S(\rho)+S(J)+S(Ab^L|J)+S(b^RC|J).
 \end{equation}
 Now, to reduce the number of POVM elements (i.e., entries of the $J$-register with
 non-zero probability amplitude),
 we invoke a theorem of Davies \cite{Davies} on extremal POVMs:
 One looks at all real vectors $(\lambda_j)_j$ such that the operators
 $\lambda_j M_j$ form a POVM, i.e., $\sum_j \lambda_j M_j = \1_B$. It is
 clear that the all-ones vector is eligible, and that this set is
 compact and convex -- in fact, it is a polytope, and Davies' theorem
 states that its extremal points have at most $d_B^2$ non-zero
 entries (actually, this is just a special case of Caratheodory's lemma).
 On the other hand, the all-ones vector can be convex-decomposed
 into extremal ones, i.e.,
 \begin{equation}
   \forall j \quad M_j = \sum_k r_k \lambda_j^{(k)} M_j,
 \end{equation}
 with extremal vectors $(\lambda_j^{(k)})_j$ and positive reals $r_k$
 with $\sum_k r_k = 1$. In operational terms, the POVM $(M_j)$ is
 equivalent to choosing $K=k$ with probability $r_k$ and then
 measuring the POVM $(\lambda_j^{(k)} M_j)$. This means that we can
 extend the state $\Omega$ above to
 \begin{equation}
   \Omega_{Ab^Lb^RCJK} = \sum_{jk} r_k
                                   \bigl( \1_{AC}\otimes\tau_j\sqrt{\lambda_j^{(k)}M_j} \bigr)
                                      \rho_{ABC}
                                   \bigl( \1_{AC}\otimes\sqrt{\lambda_j^{(k)}M_j}\tau_j^\dagger \bigr)
                                     \otimes \proj{j}_J \otimes \proj{k}_K,
 \end{equation}
 of which it can be readily verified that tracing over $K$ gives
 eq. (\ref{eq:Omega}). Then, by the concavity of the von Neumann
 entropy, $S(J) \geq S(J|K)$ and by the way we constructed the POVMs,
 \begin{equation}
   S(Ab^L|J) = S(Ab^L|JK), \qquad S(b^RC|J) = S(b^RC|JK).
 \end{equation}
 Hence, eq. (\ref{eq:D-compact}) is lower bounded by
 \begin{equation}
   -S(\rho)+S(J|K)+S(Ab^L|JK)+S(b^RC|JK),
 \end{equation}
 and there exists a value $k$ of $K$ for which
 \begin{equation}
   D(\rho\|\omega[\delta,\tau]) \geq -S(\rho)+S(J|K=k)+S(Ab^L|J\,K=k)+S(b^RC|J\,K=k).
 \end{equation}
 But for each $K=k$, the right hand side is a relative entropy with
 a Markov state referring to the POVM $(\lambda_j^{(k)} M_j)$; it can be
 lifted, by Naimark's theorem, to an orthogonal measurement on a
 larger space $\widehat{B}$.
 It is clear that w.l.o.g.~$B_j$ has dimension at most $d_B$: the state
 $\omega^{(j)}_{AB_jC}$ is supported in $B_j$ on a subspace of dimension
 at most $d_B$.

 {\bf 2.}
 Now for the second part: looking at eq. (\ref{eq:relent-oplus-ox}), we see that once $\delta$
 is fixed, we have states $\omega^{(j)}_{AB_jC}$ and we need to find, for each $j$
 individually, a decomposition/embedding
 $\tau_j:B_j\hookrightarrow b_j^L \otimes b_j^R$ that minimises the
 term $S(\sigma^{(j)}_{Ab_j^L})+S(\chi^{(j)}_{b_j^RC})$ in
 eq. (\ref{eq:relent-oplus-ox}). Dropping the index $j$ for now, since we will keep it fixed,
 let us introduce a purification $\ket{\phi}_{ABCD}$ of $\omega_{ABC}$;
 then, with the isometric embedding $\tau: B \hookrightarrow b^Lb^R$ implicit and
 the slight abuse of notation
 \begin{equation}
   \ket{\phi}_{Ab^Lb^RCD} := (\1_{ACD}\otimes\tau)\ket{\phi}_{ABCD},
 \end{equation}
 our task is to minimise, over all choices of $\tau$,
 \begin{equation}
   \label{eq:sum}
   S(Ab^L)+S(b^RC) = S(Ab^L)+S(ADb^L).
 \end{equation}
 Now notice that the latter quantity refers only to subsystems $AD$ and $b^L$,
 and that hence we can describe it entirely by the state
 $\tr_C \phi_{ABCD} =: \vartheta_{ABD}$
 and the completely positive and trace preserving map $T := \tr_{b^R} \circ \tau$
 mapping density operators on $B$ to density operators on $b^L$ -- by Stinespring's
 theorem, conversely every such quantum channel can be lifted to an isometric
 dilation $\tau:B \hookrightarrow b^L \ox b^R$ (the system $b^R$ would be called
 the environment of the channel).
 For fixed output system $b^L$ the set of these quantum channels is convex
 and the state $(\1_{AD}\otimes T)\vartheta_{ABD}$ is a linear function of the
 map. Hence, by the concavity of the von Neumann entropy $S$, the smallest
 sum of entropies (\ref{eq:sum}) is attained for extremal channels, which
 by a theorem of Choi~\cite{choi} have at most $d_B$ operator terms in the
 Kraus decomposition -- which translates into a dimension of at most $d_B$
 of $b^R$. The dimensionality of the subsystem covered by the output of
 that channel in $b^L$ is thus at most $d_B^2$. But now we can run the same
 argument for $b^R$ instead -- the whole setup is symmetric, so the channel from
 $B$ to $b^R$ is also w.l.o.g.~extremal, entailing $\dim b_L \leq d_B$
 (note that we fix the output dimension here to $\leq d_B$ from the previous
 argument).
\end{proof}

There is a special case of the second part of the above proof in the
literature that has inspired the present argument: that is the dimension bounds
in the so-called \emph{entanglement of purification}~\cite{E-puri}. There it
was shown that in the problem of, for a \emph{pure} state $\omega_{ABC}$,
minimising the entropy
\begin{equation}
  S(AE) = \frac{1}{2}\bigl( S(AE)+S(CF) \bigr),
\end{equation}
over all isometric embeddings $B\hookrightarrow EF$, one may restrict to
a priori bounded dimensions $\dim E = d_B$ and $\dim F = d_B^2$,
or, vice versa, $\dim E = d_B^2$ and $\dim F = d_B$. What is noticed above
is that, apart from the generalisation to mixed states, one can apply
the argument of the extremal channels twice, to get the same bound $d_B$
on the dimensions of both $E$ and $F$:

\begin{corollary}
  The entanglement of purification,
  \begin{equation}
    E_P(\rho_{AC}) = \inf_{B\hookrightarrow EF} S(AE),
  \end{equation}
  the entropy understood with respect to the state $\phi_{AEFC}$,
  is attained at an embedding with dimensions
  $\dim E,\dim F \leq d_B = \operatorname{rank}\, \rho_{AC}$. \qed
\end{corollary}

\begin{theorem}
\label{thm:lowerbound}
For any state $\rho_{ABC}$, the quantity $\Delta(\rho_{ABC})$ has
the following lower bound:
\begin{equation}
  \label{eq:lower}
  \Delta(\rho) \geq I(A:C|B)_\rho.
\end{equation}
\end{theorem}
\begin{proof}
Indeed, it is sufficient to show, for any $\rho_{ABC}$ and
decomposition of $B$ as in eq. (12) with accompanying state
$\omega[\delta,\tau]$, that
\begin{equation}
  D\bigl( \rho \| \omega[\delta,\tau] \bigr) \geq I(A:C|B)_\rho,
\end{equation}
which, by eq. (17), is equivalent to
\begin{equation}
  H(\underline{q})
   + \sum_j q_j \bigl( S(\sigma^{(j)}_{Ab_j^L}) + S(\chi^{(j)}_{b_j^RC}) \bigr)
                                          \geq S(B)_\rho + S(A|B)_\rho + S(C|B)_\rho.
\end{equation}
It turns out to be convenient to introduce the following state of
five registers to represent the entropic quantities in the above:
\begin{equation}
  \Omega_{J A b^L b^R C} = \sum_{j} q_j \ketbra{j}{j}_J \otimes \omega^{(j)}_{Ab_j^Lb_j^RC},
\end{equation}
observing that we may think of all $b_j^{L}$, $b_j^{R}$ as subspaces of one
$b^L$, $b^R$, respectively. Then the inequality we need to prove reads
\begin{equation}
  S(J)_\Omega + S(Ab^L|J)_\Omega + S(b^RC|J)_\Omega
                                          \geq S(B)_\rho + S(A|B)_\rho + S(C|B)_\rho.
\end{equation}
This is done by invoking standard inequalities as follows:
\begin{equation}\begin{split}
  S(J)_\Omega + S(Ab^L|J)_\Omega + S(b^RC|J)_\Omega
          &=    S(J)_\Omega + S(A|b^LJ)_\Omega + S(b^L|J)_\Omega
                            + S(C|b^RJ)_\Omega + S(b^R|J)_\Omega                       \\
          &\geq S(J)_\Omega + S(b^Lb^R|J)_\Omega + S(A|b^LJ)_\Omega + S(C|b^RJ)_\Omega \\
          &=    S(Jb^Lb^R)_\Omega + S(A|b^LJ)_\Omega + S(C|b^RJ)_\Omega                \\
          &\geq S(B)_\rho         + S(A|B)_\rho      + S(C|B)_\rho,
\end{split}\end{equation}
where in the second line we have used ordinary subadditivity of
entropy, and in the fourth line the fact that $\Omega$ is obtained
from $\rho$ by a unital c.p.t.p. map on $B$; it can only increase
the entropy, and, since it induces c.p.t.p. maps from $B$ to
$Jb^L$ and $Jb^R$, we can use the non-decrease of the conditional entropy
under processing of the condition (that's basically strong
subadditivity).
\end{proof}

\medskip
That means, for given dimensions $d_A$, $d_B$,
$d_C$, we may define the continuous and monotonic real function
\begin{equation}
  \Delta(t;d_A,d_B,d_C) := \max\bigl\{ \Delta(\rho_{ABC})
                                           \,:\, I(A:C|B)_\rho \leq t \bigr\},
\end{equation}
which has the property $\Delta(t;d_A,d_B,d_C) = 0$ if and only if $t=0$ and
$\Delta(t;d_A,d_B,d_C) \geq t$ (for not too large $t$, i.e.~$t\leq 2
\log{\min{\{d_A,d_C}\}}$).

\section{Pure states}
\label{sec:pure}
Here we give some results when $\rho$ is a pure state.
The entropy of the density matrix of a pure state is zero,
and the minimum over $\tau_j$ of the $j^{\text{th}}$ von Neumann
entropy term in the sum in eq.~(\ref{eq:relent-oplus-ox}) is the
entanglement of purification of $\rho_{AC}^{(j)}$,~\cite{E-puri}.
Thus, we arrive at the formula
\begin{equation}
  \label{eq:pure-D}
  \min_{\tau} D\bigl( \psi\|\omega[\delta,\tau] \bigr)
                = H(\underline{q}) + 2 \sum_j q_j E_P\bigl( \rho_{AC}^{(j)} \bigr).
\end{equation}
Using the calculation of $E_P$ for symmetric and antisymmetric
states in~\cite{CW05}, we can now show:
\begin{theorem}
  \label{thm:pure-D-bounds}
  Let $A$ and $C$ be systems of the same dimension $d$.
  For any pure state $\psi_{ABC}$ such that $\rho_{AC} = \tr_B \psi_{ABC}$
  is supported either on the symmetric or on the antisymmetric subspace
  of $AC$, we have
  \begin{equation}
    S(\rho_A) \leq \Delta(\psi) \leq 2 S(\rho_A).
  \end{equation}
\end{theorem}
\begin{proof}
The upper bound can be simply derived by considering a single
decomposition of system $B$ and calculating the value of
$D(\psi\|\omega[\delta,\tau])$. Since $\Delta(\rho)$ is a minimum
over all possible decompositions of system $B$, choosing one will
immediately give an upper bound. Consider the following
decomposition,
\begin{equation}
B = b^L \otimes b^R := B \otimes \CC.
\end{equation}
This gives a single term of tensor products leading to the following
density matrix
\begin{equation}
\omega[\delta,\tau]=\rho_{AB} \ox \rho_C,
\end{equation}
therefore we have,
\begin{equation}
D(\psi\|\omega[\delta,\tau])=2E_P(\rho_{AC}).
\end{equation}
A property of the entanglement of purification~\cite{E-puri} is
that if a two-party state $\rho_{AC}$ is completely supported
either on the symmetric or antisymmetric subspace of $AC$ then
the entanglement of purification is simply the entropy of reduced
state of one of the parties~\cite{CW05},
\begin{equation}
\label{eq:eopsas}
 E_P(\rho_{AC})=S(\rho_A)=S(\rho_C).
\end{equation}
Hence we prove the upper bound.

The lower bound is a consequence of strong subadditivity of
quantum entropy. We know from eq. (\ref{eq:eopsas}) that
\begin{equation}
\Delta(\rho)=H(\underline{q})+2\sum_jq_jS(\rho_A^{(j)}) \geq
H(\underline{q})+\sum_jq_jS(\rho_A^{(j)}).
\end{equation}
Note, however that
\begin{equation}
H(\underline{q})+\sum_jq_jS(\rho_A^{(j)})\geq S\bigg(\sum_j
q_j\rho_A^{(j)}\bigg) = S(\rho_A).
\end{equation}
Hence we have shown the lower bound and this concludes the proof.
\end{proof}

\section{Examples}
\label{sec:examples}
In this section we examine families of states
which we can use to numerically illustrate the bounds on
$\Delta(\rho)$. We look at two families of examples: first, on
three qubits,
\begin{example}
{\normalfont
Consider the following family of three qubit states
\begin{equation}
  \ket{\psi(x)}_{ABC}
           := \frac{1}{\sqrt{2}}
                \bigl( \ket{\varphi_x}_A\ket{0}_B\ket{\varphi_x}_C
                     + \ket{\varphi_{-x}}_A\ket{1}_B\ket{\varphi_{-x}}_C \bigr),
\end{equation}
where $\ket{\varphi_x} := \sqrt{1-x^2}\ket{0}+x\ket{1}$, and $x$
is a real parameter. Using the  notation $y=\sqrt{1-x^2}$ so that
$y^2+x^2=1$, we can calculate the following reduced density
matrices for this pure state:
\begin{equation}
\rho_A = \rho_C = \left(%
\begin{array}{cc}
  y^2 & 0 \\
  0 & x^2 \\
\end{array}%
\right),
\end{equation}
\begin{equation}
\rho_B = \frac{1}{2}
\left(%
\begin{array}{cc}
  1 & (y^2-x^2)^2 \\
  (y^2-x^2)^2 & 1 \\
\end{array}%
\right).
\end{equation}
Therefore we can calculate the entropy of each single party
density matrix.
\begin{align}
  S(\rho_A) &= S(\rho_C) = -y^2 \log{y^2} - x^2 \log{x^2} = H_2(x^2), \\
  S(\rho_B) &= -(y^4+x^4) \log{(y^4+x^4)} - 2x^2y^2 \log{2x^2y^2}.
\end{align}
From theorem~\ref{thm:pure-D-bounds} we know that for totally
symmetric or totally anti-symmetric states, $S(\rho_A)\leq
\Delta(\rho)\leq 2S(\rho_A)$. Note also that for this state
$I(A:C|B)_{\psi(x)} = S(AB)+S(BC)-S(B) = 2S(A)-S(B)$. Thus, we
wish to understand the ratio
\begin{equation}
\frac{S(\rho_A)}{2S(\rho_A)-S(\rho_B)}.
\end{equation}
If we look at the leading order terms of the single party entropies,
since $0<x<1$, we know that $x^2 \log{x}$ and $x^2$ are of
lower order than $x^4$, $x^6$, etc. Thus, only taking $x^2\log x$
and $x^2$ terms,
\begin{align}
 S(\rho_A) &= -(1-x^2)\log{(1-x^2)} - 2x^2 \log{x}                             \nonumber\\
           &= - \frac{2x^2 \ln{x}}{\ln{2}} + \frac{x^2}{\ln 2} + O(x^4),       \\
 S(\rho_B) &= -(1-2x^2+2x^4)\log(1-2x^2+2x^4)                                  \nonumber\\
           &\phantom{=}
              - 2x^2(1-x^2) [1 + 2\log{x} + \log(1-x^2)]                       \nonumber\\
           &= - \frac{4x^2\ln{x}}{\ln{2}} + \frac{2x^2}{\ln{2}} - 2x^2 + O(x^4).
\end{align}
Inserting these expressions, we find.
\begin{equation}
  \frac{\Delta(\rho)}{I(A:C|B)_\rho}
       \geq  \frac{S(\rho_A)}{2S(\rho_A)-S(\rho_B)}
       =     -\frac{\ln 2}{\ln x} + O(1)\quad {\rm as}\ x\rightarrow 0.
\end{equation}
Therefore for this state we can make this quantity approach
$+\infty$ as the value of $x$ decreases, marking a striking
deviation from the classical case.
}
\end{example}

\begin{example}
\label{example:d}
{\normalfont
Another use of theorem~\ref{thm:pure-D-bounds} is for the pure states
$\ket{\zeta(d)}_{ABC}$ on systems $A$ and $C$ of dimension $d$ and
$B$ of dimension $d(d+1)/2$: namely, $\ket{\zeta(d)}$ is the purification
of the completely mixed state on the symmetric subspace of $AC$,
i.e.~$\zeta_{AC} = \tr_B \zeta_{ABC}$ is proportional to the
symmetric subspace projector, of rank $d(d+1)/2$. For this family
of states, we have
\begin{equation}
  I(A:C|B)_{\zeta(d)} = 1 + \log\frac{d}{d+1} < 1,
\end{equation}
while according to our theorem,
\begin{equation}
  \Delta\bigl( \zeta(d) \bigr) \geq S(A) = \log d.
\end{equation}
This example shows that not only must any bound on $\Delta$ depend nonlinearly
on $I(A:C|B)$, but that a $\log$-dimensional factor is also necessary.
}
\end{example}

\begin{example}
{\normalfont
Consider the class of states
\begin{equation}
  \rho_{ABC} = \sum_j p_j \proj{j}_A \otimes \proj{\psi_j}_B \otimes \proj{j}_C,
\end{equation}
characterised by an ensemble of pure states $\{ p_j, \ket{\psi_j} \}$
on $B$ -- the states of $A$ and $C$ are meant to be mutually
orthogonal states. For a POVM $(M_k)$ on $B$, and using the
previous notation of $\widehat{B} = K b^L b^R$, the optimal
state is given by
\begin{equation}
  \omega[\delta,\tau]_{A\widehat{B}C}
      = \sum_{jk} p_j \proj{j}_A \otimes
                      \bigl( \sqrt{M_k}\proj{\psi_j}\sqrt{M_k} \bigr)_{b^Lb^R}
                      \otimes \proj{k}_K \otimes \proj{j}_C.
\end{equation}
We can calculate the following using formula (\ref{eq:D-compact})
using the fact the system is symmetric in systems $A$ and $C$ and
$S(\rho)=S(A)$,
\begin{equation}
  D(\rho\|\omega[\delta,\tau]) = -S(A) + S(K) +  S(Ab^L|K) +
  S(Ab^R|K)
\end{equation}
It is fairly clear from the formula since $S(Ab^L|K)+S(Ab^R|K)
\geq 2S(A|K)$ that the optimal choice of $b^Lb^R$ is to make one
trivial, the other $B$, so that
\begin{equation}
  D(\rho\|\omega[\delta,\tau]) = -S(A) + S(K) + 2 S(A|K)
                               = S(A|K) + S(K|A),
\end{equation}
all entropies relative to the state $\omega$. Note that $A$ and
$K$ are essentially classical registers, so that the above is
really a classical probabilistic/entropic formula for the relative
entropy. It is also quite amusing to see a quantity appearing
that is known as information-distance in other contexts (see
e.g.~\cite{Vitanyi}).
}
\end{example}

\section{Conclusions}
\label{sec:conclusion} We have investigated the relation between
the quantum conditional mutual information of a three-party state,
and its relative entropy distance from the set of all (short)
quantum Markov chains. While the latter is always larger or equal
than the former, with equality in the classical case, in general
the relative entropy distance can be much larger than the
conditional mutual information. We showed this by developing tools
to lower bound the relative entropy distance, in particular for
pure states of a special symmetric form. In the process we found
many useful properties of the minimum relative entropy distance
from Markov states. Our findings indicate that the
characterisation of quantum Markov chains in terms of vanishing
quantum conditional mutual information is not robust, or at least
not at all like the classical case, or the (quantum and classical)
case of ordinary mutual information. Since these lower bounds are
additive for tensor products of states,  this surprising and
perhaps displeasing behaviour will not go away in an asymptotic
limit of many copies of the state.

What we haven't found is an \emph{upper bound} of the relative
entropy distance $\Delta$ in terms of the conditional
mutual information $I(A:C|B)$; our examples above show that
such a bound has to depend nonlinearly on $I$ and it has to
contain a factor proportional to the logarithm of one or more
of the local dimensions. Note that if there were a bound of the
form $\Delta(\rho) \leq f(I)\log(d_A d_C)$ -- in particular
not depending on the dimension of $B$ --, then this would
settle a question left open in~\cite{CW04}: namely, it would imply
that the ``squashed entanglement'' $E_{\rm sq}(\rho_{AB})$
of a bipartite state $\rho_{AB}$ is zero if and only if the state
is separable. (We are grateful to Pawe\l{} Horodecki for pointing
this out to us.)

We close by pointing out that our results cast doubts on
earlier ideas of two of the present authors (NL and AW),
reported in~\cite{Linden:Winter}, on how to
prove a non-standard inequality for the von Neumann entropy.
The heuristics given there don't seem to bear out, in the
light of the present paper; of course, the conjectured entropy
inequality itself may well still be true.

\acknowledgments
It is our pleasure to acknowledge discussions on the topics of this
paper with Micha\l{} and Pawe\l{} Horodecki.

BI, NL and AW acknowledge
support from the U.K.~Engineering and Physical Sciences Research
Council through ``QIP IRC''; NL and AW furthermore were supported
through the EC project QAP
(contract IST-2005-15848), and AW gratefully acknowledges support via a
University of Bristol Research Fellowship.

\end{document}